\documentclass[a4paper,american, 12pt]{scrartcl}

\makeatletter
\newcommand\iraggedright{%
	\let\\\@centercr\@rightskip\@flushglue \rightskip\@rightskip
	\leftskip\z@skip}
\makeatother
\iraggedright

\usepackage{tikz-cd}
\usepackage{verbatim}
\usepackage{enumerate}
\usepackage{lmodern}
\usepackage[utf8]{inputenc}
\usepackage[T1]{fontenc}
\usepackage{color}
\usepackage{babel}
\usepackage{amsfonts, amsmath, amsthm, amssymb}
\usepackage{setspace}
\usepackage[authoryear]{natbib}
\usepackage[font=small]{quoting}
\usepackage{latexsym}
\usepackage[bookmarks=false,pdfborder={0 0 0},colorlinks=false]{hyperref}
\usepackage{dsfont}
\usepackage{xcolor}

\makeatletter

\usepackage[arrow, matrix, curve]{xy}

\theoremstyle{plain}

\theoremstyle{definition}

\title{On Boltzmann vs. Gibbs and the Equilibrium in Statistical Mechanics} 
\author{Dustin Lazarovici\thanks{Dustin.Lazarovici@unil.ch} \\ Université de Lausanne, Faculté des Lettres,\\
	Section de Philosophie, 1015 Lausanne, Switzerland}

\begin{document}
	\maketitle

\begin{abstract}
	\noindent In a recent article, Werndl and Frigg discuss the relationship between the Boltzmannian and Gibbsian framework of statistical mechanics, addressing in particular the question when equilibrium values calculated in both frameworks agree. In this paper, I address conceptual confusions that could arise from their discussion, concerning in particular the authors' use of ``Boltzmann equilibrium''. I also clarify the status of the Khinchin condition for the equivalence of Boltzmannian and Gibbsian, and show that it follows under the assumptions proposed by Werndl and Frigg from standard arguments in probability theory. 
\end{abstract}
\section{Boltzmann vs. Gibbs}
\noindent In a recent paper,
Charlotte Werndl and Roman Frigg (2017) \nocite{werndl.frigg2017} discuss the relationship between the Boltzmannian and Gibbsian framework of statistical mechanics, addressing in particular the question, if and when the two formulations yield equivalent predictions for equilibrium values of macroscopic observables defined in terms of appropriate macro-variables. 

In the Boltzmannian framework, the macro-variables take (approximately) constant values on the equilibrium region of phase space which are thus revealed by a suitable measurement on a system in equilibrium (a system, that is, whose actual micro-state is in the equilibrium region). The role of the stationary phase space measure is to establish that it is very likely to find the system in an equilibrium state. 
In the Gibbsian framework, equilibrium is a property of an \emph{ensemble}, represented by a stationary distribution $\rho$ on phase space $\Gamma$, and it is often (though maybe somewhat carelessly) said that the prediction for a measurement of a macro-variable $f$ on an individual ensemble system is given by the \emph{phase average}
\begin{equation} \label{phase}\langle f \rangle = \int_{\Gamma} f(x) \rho(x) \, \mathrm{d}x\, , \end{equation} where $x\in \Gamma$ is the phase space variable. This quantity is also called the \emph{ensemble average} or \emph{expectation value} of $f$. 


There are many situations in which the Gibbsian phase average agrees -- within appropriate error bounds -- with the Boltzmannian equilibrium value. Werndl and Frigg mention a criterion which they call the ``Khinchin condition'' and which they characterize briefly as the phase function having ``small dispersion for systems with a large number of constituents''. Indeed, in less technical terms, a sufficiently small dispersion of the macro-variable means precisely that typical values (the Boltzmann equilibrium value) are close to the average value (the Gibbsian equilibrium value). Another way to formulate the Khninchin condition -- now from a Boltzmannian perspective -- is to say that there exists a unique Boltzmann equilibrium whose corresponding macro-region exhausts almost the entire phase space volume. Formally: 

\begin{equation}\label{Khinchin} \mu_\rho\left(\Gamma_{\mathrm{eq}} \right) = \int_\Gamma \mathds{1}\left(\left\lbrace f(x) \in \left( \xi-\Delta\xi, \xi +\Delta\xi \right)\right\rbrace \right) \rho(x) \, \mathrm{d}x= 1 -\epsilon, \end{equation}
where $\Delta\xi$ is very small compared to $\xi$ and $\epsilon$ is very small compared to $1$ (and $\mathds{1}(\cdot)$ is the characteristic function). For then, the macro-variable $f$ takes an (approximately) constant value -- the Boltzmannian equilibrium value $\xi \pm \Delta\xi$ -- on a set of measure close to $1$ -- the Boltzmannian equilibrium region $\Gamma_{\mathrm{eq}}$. Hence, the phase average \eqref{phase} will be close to the Boltzmannian equilibrium value, provided $f$ is somewhat well-behaved and its values don't suddenly ``explode'' outside the equilibrium region. Rigorously:  
\begin{equation}\begin{split}\label{Khinchinestimate}
\lvert \langle f \rangle - (1-\epsilon)\xi\rvert  &\leq 
 \int_{\Gamma_{\mathrm{eq}}} \lvert f(x) - \xi \rvert\, \rho(x) \, \mathrm{d}x + \int_{\Gamma\setminus\Gamma_{\mathrm{eq}}} \lvert f(x) \rvert\rho(x) \, \mathrm{d}x \\
 &\leq (1-\epsilon) \Delta\xi + \epsilon \sup_{x \in \Gamma\setminus\Gamma_{\mathrm{eq}}} \lvert f(x)\rvert,
\end{split}\end{equation} 
and thus $ \langle f \rangle \approx \xi  $ if $\epsilon \sup_{x \in \Gamma\setminus\Gamma_{\mathrm{eq}}} \lvert f(x)\rvert \ll \lvert \xi\rvert$, i.e. unless the non-equilibrium values of $f$ are many orders of magnitude larger than the equilibrium value.

It is important to emphasize that (unless one considers a thermodynamic limit) ``the Boltzmannian equilibrium value''  refers in general to a small range of values of $f$. This kind of coarse-graining is both essential for probabilistic estimates and physically called for, since the relevant measurements have limited accuracy. For Werndl and Frigg, in contrast, each single value of the macro-variable $f$ defines a different ``macro-state'', implying that, for them, even the slightest variation in the respective physical quantity -- pressure, density, energy etc. -- leads to a system being ``out of equilibrium''. What the authors call the ``Boltzmann equilibrum'' is thus not the equilibrium as defined by Boltzmann or used in Boltzmannian statistical mechanics, and it is unfortunate, since miselading, that they refer to it by the same name. For our futher discussion, we will thus refer to it as the ``Werndl-Frigg equilibrium'' instead.  

The existence of a dominant Boltzmann equilibrium (in the sense of Boltzmann) is in fact the \emph{generic} case in statistical mechanics, giving rise to the claim that Boltzmann and Gibbs make (in general) equivalent predictions for systems in the respective equilibria. If one agrees that the Boltzmannian formulation is the more fundamental one, this also explains \emph{why} Gibbsian phase averaging yields (in general) accurate predictions for individual measurements.

On the other hand, there are famous and well-studied cases in which the Khinchin condition \eqref{Khinchin} doesn't hold. For instance, in the two-dimensional Ising model without external field, it makes sense to speak of \emph{two} Boltzmann equilibria below the critical temperature, corresponding to a positive or negative magnetization, respectively. The Gibbsian distribution $\rho$ is, however, symmetric under a flip of all spins, hence yielding an average magnetization of zero. There is nothing inconsistent or mysterious about this fact, as long as one keeps in mind that the Gibbsian value refers, in the first place, to an ensemble average. In particular, in statistical mechanics, one does not try to draw interesting conclusions about the Ising model from such phase averages. Instead, one usually studies so-called phase transitions at the critical temperature by fixing either $+1$ or $-1$ boundary conditions (referring to the polarization of spins at the edge of the lattice) thus implicitly picking one of the two magnetization states.

\section{Boltzmann vs. Werndl and Frigg}

In their paper, Werndl and Frigg mention the magnetization in the Ising model only briefly, but present instead other examples for which they claim a disagreement between Boltzmannian and Gibbsian equilibrium values. One such example is based on the ``baker's gas'', a mathematical model for the ideal gas that the authors have used in various publications to argue that the Boltzmann equilibrium fails to be \emph{dominant}, i.e. exhaust a majority of the phase space volume as stated in equation \eqref{Khinchin}. To this end, the authors partition the one-particle phase space (corresponding to the unit square in the baker's model) into $k$ cells and claim that the macro-state corresponding to the Boltzmann equilibrium is the ``uniform distribution'' for which each cells contains exactly $\frac{N}{k}$ particles. It is then easy to see that while the phase space volume associated with this uniform distribution is greater than the phase space volume associated with any other particular arrangement of particles over the cells, it will not exhaust a majority of phase space volume for large $N$. In the present paper, the authors exploit this fact -- amounting to an apparent violation of the Khinchin condition \eqref{Khinchin} -- by introducing an artificial macro-variable, weighing the previously defined ``macro-regions'' in such a way that the Gibbsian phase average differs significantly from the value associated with the uniform distribution.

However, as emphasized above, the authors' reference to the ``Boltzmann equilibrium'' is a misnomer, since they use a notion of equilibrium that does not correspond to the concept introduced by Boltzmann and used in Boltzmannian statistical mechanics (thereby repeating the arguments of \cite{lavis2005} already criticised by \cite{lazarovici.reichert2015}). \emph{Exact} uniform distributions, where each cell contains exactly $\frac{N}{k}$ particles, are very special configurations, their measure actually goes to zero for large $N$. But configurations for which the density of particles in each cells differs only slightly from $\frac{1}{k}$ are macroscopically indistinguishable and coarse-grain to the same macro-state in Boltzmann's sense. Otherwise, we would have to say, for instance, that a gas is ``out of equilibrium'' if the left-hand-side of the volume contains even a single particle more than the right-hand-side.


Compare this with Boltzmann's discussions of the Maxwellian velocity distribution as the equilibrium distribution of an ideal gas. Here, Boltzmann was very explicit about the fact that ``for a finite number of molecules, the Maxwell distribution will not hold exactly but only to a good approximation.'' \cite[translation D.L.]{boltzmann1896} 

In the case of the baker's model, the dominant Boltzmann equilibrium with respect to the considered uniform measure contains all configurations for which the relative number of particles in each cell is within $\frac{1}{k} \pm \frac{1}{\sqrt{N}}$ (I am neglecting a $k$-dependent constant that would yield somewhat better bounds). Since $\frac{1}{\sqrt{N}}$ is a tiny number for macroscopic $N$, corresponding to density fluctuations of less than one tenth of a billionth of a percent, these configurations look \emph{macroscopically} uniform and constitute the relevant equilibrium state. The Khinchin condition \eqref{Khinchin} is thus satisfied and the Boltzmannian and Gibbsian equilibrium values will be equivalent for any sensible macro-variable as specified above. 



\section{Law of large numbers vs. EET}

By the same token, all other examples presented by Werndl and Frigg may show a disagreement between the Gibbsian equilibrium and their own, but shed no light on the relation with the real Boltzmann equilibrium. Nevertheless, the authors go on to conclude that it is ``[a]n important task of the foundations of SM [statistical mechanics] ... to classify under which conditions the two frameworks lead to the same results and under which conditions they do not'' (p. 1300) and present a ``new theorem specifying a set of conditions'' under which ``Boltzmannian'' and Gibbsian equilibrium values coincide. I quote their result in full: 

\begin{quote}\textbf{Equilibrium Equivalence Theorem (EET):} Suppose that the system $(X, T_t, \mu_X)$ is
	composed of $N \geq 1$ constituents. That is, the state $x \in X$ is given by the $N$
	coordinates $x = (x_1, . . . , x_N);\, X = X_1 \times X_2 \ldots  \times X_N$, where $X_i = X_{oc}$ for all $1 \leq  i \leq N$ ($X_{oc}$ is the one-constituent space). Let $\mu_X$ be the product measure $\mu_{X_1} \times \mu_{X_2} \ldots \times \mu_{X_N}$, where $\mu_{X_i} = \mu_{X_{oc}}$
	is the measure on $X_{oc}$. Suppose that an
	observable $\kappa$ is defined on the one-particle space $X_{oc}$ and takes the values $\kappa_1, \ldots , \kappa_k$
	with equal probability $1/k, k \leq N$. Suppose that the macro-variable $K$ is the sum
	of the one-component observable, i.e. $K(x) = \sum_{i=1}^N \kappa(x_i)$. Then the value
	corresponding to the largest macro-region as well as the value obtained by phase
	space averaging is $\frac{N}{k}
	(\kappa_1 + \kappa_2 + \ldots \kappa_k)$.
\end{quote}

\noindent This theorem tries to identify the Werndl-Frigg equilibrium value with the Gibbs average. It does not relate to the Boltzmann equilibrium. That is because the ``largest macro-region'' of Werndl and Frigg refers to the set on which the macro-variable takes \emph{exactly} the average value, and this is, again, in contrast to the Boltzmannian framework, in which a small range of (for all practical purposes indistinguishable) values coarse-grains to the same macro-state. 

It should be noted, however, that part of the conditions specified by Werndl and Frigg are sufficient (though by no means necessary) for the equivalence of Gibbsian and Boltzmannian equilibrium values. In fact, under these conditions, the equivalence is a standard exercise in statistical mechanics, based on the \emph{law of large numbers} (LLN) -- the fundamental theorem of probability theory. The LLN states that, for a family of independent and identically distributed random variables: 
\begin{equation}\label{LLN} \mu\left(\left\lbrace x : \left \lvert \frac{1}{N} \sum_{i=1}^N \kappa(x_i) - \frac{1}{k} (\kappa_1 + \kappa_2 + \ldots +\kappa_k) \right \rvert < \epsilon \right\rbrace \right) \geq 1- \frac{\sigma^2}{\epsilon^2 N} \,, \end{equation} 
for any $\epsilon >0$, where $\sigma^2$ is the variance of $\kappa$. For the macro-variable $K(x) = \sum_{i=1}^N \kappa(x_i)$ then, which is extensive and growing with $N$, we can set $\epsilon = N^{-\delta}$ for $\delta \in [0, \frac{1}{2}]$ so that, in terms of $K$ (and writing $\overline{K}:=\frac{N}{k} (\kappa_1 + \kappa_2 + \ldots +\kappa_k)$ as a ``phase average''), the LLN estimate becomes
\begin{equation}\label{LLN2} \mu\left(\left\lbrace x : \left \lvert K(x) - \int K(x')\, \mathrm{d}\mu(x') \right \rvert < N^{1-\delta} \right\rbrace \right) \geq 1- \frac{\sigma^2}{ N^{1-2\delta}} \,. \end{equation} 
Note that the bound $N^{1-\delta}$ is small compared to $\overline{K}$, which is of order $N$, i.e. it is the \emph{relative} deviation $\left\lvert \frac{K-\overline{K}}{\overline{K}} \right\rvert \lesssim \frac{N^{1-\delta}}{N} = N^{-\delta}$ that becomes vanishingly small for large particle numbers. In particular, for a macroscopic system, we have $N \sim 10^{24}$ (from Avogadro's constant), and setting $\delta = \frac{1}{3}$, we can conclude that $K$ deviates from $\overline{K}$ by less than one millionth of a percent on a set of measure (approximately) $0,999999$.\footnote{A tacit assumption, generally made, is that the variance $\sigma^2$ of the \emph{one-constituent variables} $\kappa_i$ is of order $1$. If $\sigma$ is extremely large, or somehow chosen to increase with $N$, the LLN may fail to provide relevant estimates, though such cases seem unphysical.}



Summing up in less technical terms, the weak law of large numbers states precisely that for large $N$ (which is the relevant case in statistical mechanics), phase space is dominated by an equilibrium region, on which the value of the macro-variable $K$ is very close to the expectation value (= phase average). And this is precisely the empirical equivalence of Boltzmannian and Gibbsian equilibrium values. The LLN also yields immediately the Khinchin condition, both in the sense of small dispersion (which is actually how the LLN is usually proven) and in the form of equation \eqref{Khinchin} (to be compared with \eqref{LLN2}).

The LLN holds, in fact, under much more general assumptions than those of the EET. In the form of eq. \eqref{LLN2}, the result holds for any sum of uncorrelated and identically distributed random variables with finite variance. This is to say that neither the uniform distribution, nor the discreteness of the random variables is required, and they don't even have to be defined on a one-constituent space. The conditions are nonetheless very strong and hardly apply to anything but idealized models. In fact, when Werndl and Frigg write: 
\begin{quote}The crucial assumptions of the theorem are (i) that
	the macro-variable is a sum of the observable on the constituent space and (ii) that the macro-variable on the constituent space corresponds to a partition with cells of equal probability. (p. 16) \end{quote}
they overlook the truly critical assumption in both their theorem and the LLN. While (i) is often justified, and the ``equal probability'' in (ii) unnecessary when one argues with the LLN, the crucial and very strong assumption is the \emph{statistical independence} of the constituent variables (their joint distribution factorizes!), which doesn't hold, in most relevant cases, due to interactions or conserved quantities. (It is precisely this loss of statistical independence that becomes significant in the Ising below the critical temperature.) Controlling correlations  -- or justifying in some other way a ``law of large numbers'' -- is in fact the key issue in many of the hard problems of statistical mechanics.


I want to emphasize again that while the LLN applies immediately under their stated assumptions, the theorem proven by Werndl and Frigg is not a LLN statement, because the authors have a different notion of ``Boltzmann equilibrium'' in mind. The LLN, as much of statistical mechanics, is about estimates. Werndl and Frigg don't do estimates. Instead, in their paper, the Werndl-Frigg equilibrium value and its equality to the phase average is supposed to be \emph{exact}. This is why, in addition to considering standard conditions for the LLN, the authors assume a particularly simple and symmetric distribution of the macro-variable for which the average coincides with the most likely value. The EET is then a simple exercise in combinatorics. It's physical relevance, however, is questionable to say the least. For one, the measure of what the authors call the ``largest macro-region'', i.e. set on which $K$ takes \emph{precisely} the value $\frac{N}{k} (\kappa_1 + \kappa_2 + \ldots +\kappa_k)$, actually goes to \emph{zero} for large $N$. Interpreting this measure probabilistically, it is thus very unlikely for a system to be in this Werndl-Frigg equilibrium. Second, while the macro-variable considered by Werndl and Frigg is discrete, different values can be very close for large $N$. Hence, many different ``macro-states'' in the sense of Werndl and Frigg will be empirically indistinguishable, given the limited resolution of measurements on macroscopic systems. This is a crucial difference between macro-states in the sense of Werndl and Frigg and macro-states in the sense of Boltzmann. Finally, it should be stressed that the relevant notion of equivalance for Boltzmannian and Gibbsian equilibrium predictions can only be \emph{empirical equivalance}, i.e. that the respective values agree to a sufficiently good approximation. 

The law of large numbers is in fact the paradigm that one should have in mind when one thinks about the Boltzmann equilibrium: there is a certain \emph{range} of typical values for the relevant quantities. And the larger the particle number $N$, the more weight (phase space measure or probability, if you wish,) is concentrated on an ever smaller range of values around the mean. 

It is also a standard result in probability theory that the variance (= dispersion squared) for a sum of independent random variables (as considered by Werndl and Frigg) is additive. This is to say, in particular, that typical fluctuations are of the order $\sqrt{N}$ and we will not have a dominant equilibrium region if the coarse-graining into macro-states is finer than that. This is a mathematical fact, not a foundational issue. (Note that partitioning the one-constituent space into cells is only a coarse-graining on the microscopic scale (order $1$ for an extensive variable) and thus not sufficient for a macroscopic coarse-graining.)




To be clear: a ``macro-variable'', qua mathematical object, is usually some nice function of the microscopic variables -- think for instance of the energy $H(q,p)$ as a function of the particles' positions and momenta in a canonical ensemble. But such a variable is in general too fine-grained to consider all its different values as macroscopically distinct. There is thus not a one-to-one correspondence between the possible values of the macro-variable and Boltzmannian macro-states. Boltzmannian macro-states are supposed to be observationally distinguishable by relevant means. 

It should also be noted that while the Boltzmannian equilibrium value is thus necessarily ``unsharp'', the Gibbsian equilibrium value, if identified with the phase average \eqref{phase}, is a definite real number by definition. It would be a mistake, however, to read this as an ``infinitely precise'' prediction of the Gibbsian theory. Instead, one can for instance compute the ensemble variance
 \begin{equation} (\Delta f)^2 :=  \label{Gibbs-error} \int (f(x)-\langle f \rangle)^2 \rho(x) \, \mathrm{d}x\, \end{equation}
and identify the Gibbsian prediction, to be compared with the Boltzmannian one, with $\langle f \rangle \pm \Delta f$.

\section{Conclusion}
In sum, I arrive at the following conclusions about the paper of Werndl and Frigg and their proposed ``Ensemble Equivalence Theorem'':\\[2ex]

\noindent 1. Contrary to what the authors announce in their paper, their theorem doesn't provide any new conditions for the agreement of Boltzmannian and Gibbsian equilibrium values. They have instead stated conditions (plus some unnecessary ones) that are standard for the LLN from which the Khinchin condition readily follows.  \\[2ex]

	
\noindent 2. The authors' suggestion that ``we need other conditions next to the Khinchin condition'' seems ill-motivated to begin with. They present examples in which what they identify as the ``Boltzmann equilibrium'' fails to be dominant -- so that the Khinchin condition \eqref{Khinchin} is not satisfied -- but those are also examples in which they find the associated equilibrium value to disagree with the Gibbsian one. In fact, a good case can be made that the Khinchin condition -- in the sense of ``uniqueness and dominance of the Boltzmann equilibrium'' -- is not only sufficient\footnote{Together with an appropriate bound on the variation of the macro-variable as specified after equation \eqref{Khinchinestimate}.} but also necessary for the empirical equivalence of Boltzmannian and Gibbsian equilibrium values. (For otherwise, we have either no Boltzmann equilibrium -- and thus no Boltzmann equilibrium value -- or multiple Boltzmann equilibria, so that the Gibbsian phase average will correspond to an average of the Boltzmann equilibrium values rather than any one in particular). However, instead of arguing this point in greater detail, a more important remark is maybe the following: If the Khinchin condition doesn't hold, it  means that there's a high probability of finding macro-values that differ significantly from the Gibbsian phase average, so that the phase average, as a prediction for individual measurements, is highly dubious in the first place.\footnote{It is common misconception that one measures ``ergodic time-averages''. This is wrong since ergodic time-scales are \emph{much} too long, see \cite{goldstein2001}.}  \\[2ex]
	

\noindent 3.  Werndl and Frigg have written a whole series of papers (2015a,b, 2017) attacking the premise of a dominant Boltzmann equilibrium and addressing the problems -- like the non-equivalance of Boltzmannian and Gibbsian equilibrium values -- that ensue\nocite{werndl.frigg2015, werndl.frigg2015a, werndl.frigg2017}. However, a reader not well-trained in statistical mechanics could hardly tell from their discussions that the foundational problems they choose to work on arise merely from their own definition of ``macro-states'' and ``Boltzmann equilibrium'' that does not correspond to the way in which physicists, since Boltzmann, have actually argued. Since the authors generally fail to consider an appropriate coarse-graining of the microscopic state space (as suggested both by physical considerations and elementary results in probability theory) their counterexamples to the existence of a dominant Werndl-Frigg equilibrium have no relevance for the Boltzmannian framework. No Boltzmannian -- least of all Boltzmann himself -- every claimed that one could partition phase space in any arbitrary matter and end up with a dominant equilibrium state. 

In general, I am skeptical of this \emph{ad absurdum} approach to the foundations of statistical mechanics -- constructing artificial counterexamples that create artificial problems. In my opinion, it misses the point of the discipline, which is not an ``axiomatic theory'' but an effective framework for the description of complex systems that requires some degree of pragmatism and good physical sense. A crucial difference is that in an axiomatic theory, any counterexample can point to foundational issues, while in statistical mechanics, such counterexamples point merely to inadequate use.

\newpage

\bibliographystyle{Chicago}
\bibliography{StatMech}

\end{document}